\documentclass{jfm}
\usepackage{graphicx}
\usepackage{amsmath}
\usepackage{natbib}
\usepackage{xr}

\begin{document}

\title[Effective slip boundary conditions for arbitrary 1D surfaces]
{Effective slip boundary conditions for arbitrary 1D surfaces}
\author[Evgeny S. Asmolov and Olga I. Vinogradova]{Evgeny S. Asmolov$%
^{1,2,3}$ \thanks{%
Author to whom correspondence should be addressed; email: aes50@yandex.ru
}
and Olga I. Vinogradova$^{1,4,5}$}

\affiliation{
$^1$A.N.~Frumkin Institute of Physical Chemistry and Electrochemistry, Russian
Academy of Sciences, 31 Leninsky Prospect, 119991 Moscow, Russia\\[\affilskip]
$^2$Central Aero-Hydrodynamic Institute, 140180
Zhukovsky, Moscow region,  Russia\\[\affilskip]
$^3$Institute of Mechanics, M. V. Lomonosov Moscow State University, 119991 Moscow,
Russia\\[\affilskip]
$^4$Department of Physics, M. V. Lomonosov Moscow State University, 119991
Moscow, Russia\\[\affilskip]
$^5$DWI, RWTH Aachen, Forckenbeckstr. 50, 52056 Aachen, Germany\\[\affilskip]}
\date{\today}
\maketitle

\begin{abstract}
In many applications it is advantageous to construct effective slip boundary
conditions, which could fully characterize flow over patterned surfaces.
Here we focus on laminar shear flows over smooth anisotropic surfaces with
arbitrary scalar slip $b(y)$, varying in only one direction. We
derive general expressions for eigenvalues of the
effective slip-length tensor, and show that the transverse component is equal
to a half of the longitudinal one with twice larger local slip, $2b(y)$. A
remarkable corollary of this relation is that the flow along any direction
of the 1D surface can be easily determined, once the longitudinal component
of the effective slip tensor is found from the known spatially nonuniform
scalar slip.
\end{abstract}

\section{Introduction}

With recent advances in microfluidics~\citep{stone2004}, renewed interest
has emerged in quantifying the effects of surface chemical heterogeneities
with a different local scalar slip on fluid motion. In this situation it is
advantageous to construct the effective slip boundary condition, which is
applied at the imaginary smooth homogeneous, but generally anisotropic
surface, and mimics the actual one along the true heterogeneously slipping
surface~\citep{vinogradova.oi:2011,Kamrin_etal:2010}. Such an effective
condition fully characterizes the flow at the real surface (on the scale
larger than the pattern characteristic length) and can be used to solve
complex hydrodynamic problems without tedious calculations.

For an anisotropic texture, the effective boundary condition generally
depends on the direction of the flow and is a tensor, $\mathbf{b}_{\mathrm{%
eff}}\equiv \{b_{ij}^{\mathrm{eff}}\}$ represented by a symmetric, positive
definite $2\times 2$ matrix~\citep{Bazant08}
\begin{equation}
\mathbf{b}_{\mathrm{eff}}=\mathbf{S}_{\theta }\left(
\begin{array}{cc}
b_{\mathrm{eff}}^{\parallel } & 0 \\
0 & b_{\mathrm{eff}}^{\perp }%
\end{array}%
\right) \mathbf{S}_{-\theta },  \label{beff_def1}
\end{equation}%
diagonalized by a rotation
\begin{equation*}
\mathbf{S}_{\theta }=\left(
\begin{array}{cc}
\cos \theta & \sin \theta \\
-\sin \theta & \cos \theta%
\end{array}%
\right) .
\end{equation*}%
Therefore, Eq.(\ref{beff_def1}) allows us to calculate an effective slip in
any direction given by an angle $\theta $. In other words, the general
problem reduces to computing the two eigenvalues, $b_{\mathrm{eff}%
}^{\parallel }$ ($\theta =0$) and $b_{\mathrm{eff}}^{\perp }$ ($\theta =\pi
/2$), which attain the maximal and minimal directional slip lengths,
respectively. This tensorial slip approach, based on a consideration of a
`macroscale' fluid motion instead of solving hydrodynamic equations at the
scale of the individual pattern, was supported by statistical diffusion
arguments~\citep{Bazant08}, and was recently justified for the case of
Stokes flow over a broad class of periodic surfaces~\citep{Kamrin_etal:2010}.

The concept of an effective tensorial slip has recently been proven by
simulations~\citep{priezjev.n:2011,harting.j:2012}, and was already used to obtain
simple solutions of several complex problems. It may be useful in many situations,
such as drainage of thin films~\citep{belyaev.av:2010b,asmolov_etal:2011},
mixing in superhydrophobic channels~\citep{vinogradova.oi:2011}, and
electrokinetics of patterned surfaces~\citep{bahga:2009,belyaev.av:2011a}.
However, to the best of our knowledge all these analytical solutions were
obtained for a flow on alternating slip and no-slip stripes, and the
quantitative understanding of effective slippage past other types of
anisotropic surfaces is still challenging.

In this paper, we study the Stokes flow past flat surfaces, where the local
(scalar) slip length $b$ varies only in one direction. Our focus is on the
limit of a thick (compared to texture period) channel or a single interface,
so that effective slip is a characteristics of a heterogeneous interface
solely and does not depend on the channel thickness. We derive a simple
universal relationship between eigenvalues of the slip-length tensor. This
allows one to avoid tedious calculations of flows in a transverse
configuration, by reducing the problem to an analysis of longitudinal flows,
which is much easier to evaluate. Our results open a possibility to solve a
broad class of hydrodynamic problems for 1D textured surfaces.

\section{Theory}

\subsection{General consideration}

\begin{figure}
\begin{center}
\vspace{0.1in} \includegraphics[width=2.5in]{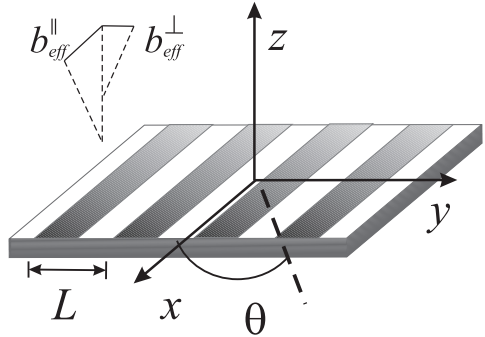}
\end{center}
\caption{Schematic representation of periodic textures with scalar slip
boundary conditions, varying in direction $y$. The patterns of slip boundary
conditions are depicted as alternating stripes with piecewise constant slip
lengths, but our discussion is more general and applies to any 1D
distribution of a local slip (e.g., sinusoidal, trapezoidal, and more). }
\label{fig:slab}
\end{figure}

We consider a creeping flow along a plane anisotropic wall, and a Cartesian
coordinate system $(x,y,z)$ (Fig. \ref{fig:slab}). The origin of
coordinates is placed at the flat interface, characterized by a slip length $%
b(y)$, spatially varying in one direction, and the texture varies over a
period $L$. Our analysis is based on the limit of a thick channel or a
single interface, so that the velocity profile sufficiently far above the
surface, at a height may be considered as a linear shear flow. Note that our
results do not apply to a thin or an arbitrary channel situation, where the
effective slip scales with the channel width~%
\citep{feuillebois.f:2009,harting.j:2012}.

Dimensionless variables are defined by using $L$ as a reference length
scale, the shear rate sufficiently far above the surface, $G,$ and the fluid
kinematic viscosity, $\nu .$ We seek the solution for the velocity profile
in the form%
\begin{equation*}
\mathbf{v}=\mathbf{U}+\mathbf{u}_{1},
\end{equation*}%
where $\mathbf{U}=z\mathbf{e}_{l},\ l=x,y,\ $is the undisturbed linear shear
flow, $\mathbf{e}_{l}$ are the unit vectors. The perturbation of the flow, $%
\mathbf{u}_{1}=\left( u,v,w\right) ,$ which is caused by the presence of the
texture and decays far from the surface at small Reynolds number $%
Re=GL^{2}/\nu $ satisfies dimensionless Stokes equations,%
\begin{gather}
\mathbf{\nabla }\cdot \mathbf{u}_{1}=0,  \label{Se} \\
\mathbf{\nabla }p-\Delta \mathbf{u}_{1}=0,  \notag
\end{gather}%
where $p$ is pressure. The boundary conditions at the wall and at infinity
are defined in the usual way%
\begin{gather}
z=0:\quad \mathbf{u}_{1\tau }-\beta \left( y\right) \frac{\partial \mathbf{u}%
_{1\tau }}{\partial z}=\beta \left( y\right) \mathbf{e}_{l} ,\   \label{bcu} \\
w=0,  \label{bcw0}
\end{gather}%
\begin{equation}
z\rightarrow \infty :\quad \frac{\partial \mathbf{u}_{1}}{\partial z}=%
\mathbf{0,}  \label{bci}
\end{equation}%
where $\mathbf{u}_{1\tau }=\left( u,v,0\right) $ is the velocity along the
wall and $\beta =b/L$ is the normalized slip length.

A local slip length can be expanded in a Fourier series%
\begin{equation*}
\beta \left( y\right) =\sum_{n=-\infty }^{\infty }b^{\ast }\left( n\right)
\exp \left( \mathrm{i}k_{n}y\right) ,
\end{equation*}%
\begin{equation*}
k_{n}=2\pi n.
\end{equation*}%
Similarly, the solution to (\ref{Se}) for $\mathbf{u}_{1}$ and $p$ has the
form
\begin{equation}
\mathbf{u}_{1}=\sum_{n=-\infty }^{\infty }\mathbf{u}^{\ast }\left(
n,z\right) \exp \left( \mathrm{i}k_{n}y\right) ,\quad p=\sum_{n=-\infty
}^{\infty }p^{\ast }\left( n,z\right) \exp \left( \mathrm{i}k_{n}y\right) ,
\label{fu}
\end{equation}%
and the Stokes equations can then be rewritten as
\begin{gather}
\mathbf{\nabla }^{\ast }\cdot \mathbf{u}^{\ast }=0,  \label{1.5} \\
\mathbf{\nabla }^{\ast }p^{\ast }-\Delta ^{\ast }\mathbf{u}^{\ast }=\mathbf{0},
\label{1.5b} \\
\mathbf{\nabla }^{\ast }=\left( 0,\mathrm{i}k_{n},\frac{\mathrm{d}}{\mathrm{d%
}z}\right) ,\quad \Delta ^{\ast }=\frac{\mathrm{d}^{2}}{\mathrm{d}z^{2}}%
-k_{n}^{2}.  \notag
\end{gather}%
We have therefore reduced the problem to the system of ordinary differential
equations (ODE), which can be now solved analytically. Similar strategy has been used before to address different  hydrodynamic problems \citep{asmolov:2008,Kamrin_etal:2010}.

The zero mode solution represents a constant, $\mathbf{u}^{\ast }\left(
0,z\right) =\left( c_{x}\left( 0\right) ,c_{y}\left( 0\right) ,0\right) .$
The eigenvalues of the effective slip-length tensor can be obtained as the
components of $\mathbf{u}^{\ast }\left( 0,z\right) :$
\begin{equation}
b_{\mathrm{eff}}^{\parallel }=Lc_{x}\left( 0\right) ,\quad b_{\mathrm{eff}%
}^{\perp }=Lc_{y}\left( 0\right) .  \label{co_b}
\end{equation}

Below we analyze in more details two configurations of the shear flow, where
the slip regions are distributed parallel ($\theta=0$) and transverse ($%
\theta=\pi/2$) to the shear flow direction.

\subsection{Longitudinal configuration}

For longitudinal patterns, $\mathbf{U}=z\mathbf{e}_{x},$ so that the
perturbation of the velocity has the only one component, $\mathbf{u}%
_{1}=\left( u,0,0\right) ,$ and the system (\ref{1.5})-(\ref{1.5b}) reduces
to a single ODE:%
\begin{equation}
\Delta ^{\ast }u^{\ast }=0.  \label{1.6}
\end{equation}%
Its decaying at infinity solution for non-zero modes has the form
\begin{equation}
u^{\ast }=c_{x}\left( n\right) \exp \left( -\left\vert k_{n}\right\vert
z\right) .  \label{1x}
\end{equation}%
The boundary condition at the wall, (\ref{bcu}), then determines constants $%
c_{x}\left( n\right) $. Indeed, from (\ref{fu}) and (\ref{1x}) we get%
\begin{equation*}
z=0:\quad u=\sum_{n=-\infty }^{\infty }c_{x}\left( n\right) \exp \left(
\mathrm{i}k_{n}y\right) ,\quad \frac{\partial u}{\partial z}%
=-\sum_{n=-\infty }^{\infty }\left\vert k_{n}\right\vert c_{x}\left(
n\right) \exp \left( \mathrm{i}k_{n}y\right) ,
\end{equation*}%
so that (\ref{bcu}) takes the form
\begin{equation}
c_{x}\left( n\right) +\sum_{m=-\infty }^{\infty }\left\vert k_{m}\right\vert
c_{x}\left( m\right) b^{\ast }\left( n-m\right) =b^{\ast }\left( n\right) ,
\label{Fbx}
\end{equation}%
where $b^{\ast }\left( n\right) $ is the Fourier coefficient of the slip
length. Thus we reduce the longitudinal problem to the infinite linear
system for $c_{x}\left( n\right) ,$ which can be found by truncating the
system and by using standard routines for linear systems.

\subsection{Transverse configuration}

For transverse patterns, $\mathbf{U}=z\mathbf{e}_{y}$ and $\mathbf{u}%
_{1}=\left( 0,v,w\right) ,$ the solution is usually constructed using the
vorticity $\mathbf{\omega =\nabla \times u}_{1}=\left( 0,0,\omega
_{z}\right) $, which significantly complicates the analysis as compared to a
longitudinal case \citep{cottin:2004,priezjev.nv:2005,Belyaev:2010}.
However, simple analytical solutions can be obtained directly for the
Fourier coefficients of velocities \citep{asmolov:2008,Kamrin_etal:2010}.
Indeed, equations (\ref{1.5})-(\ref{1.5b}) for transverse stripes can be
written as%
\begin{gather}
\mathrm{i}k_{n}v^{\ast }+\frac{\mathrm{d}w^{\ast }}{\mathrm{d}z}=0,  \notag
\\
\mathrm{i}k_{n}p^{\ast }-\Delta ^{\ast }v^{\ast }=0,  \label{2b} \\
\frac{\mathrm{d}p^{\ast }}{\mathrm{d}z}-\Delta ^{\ast }w^{\ast }=0.  \notag
\end{gather}%
By excluding $p^{\ast }\ $and $v^{\ast },$ one can transforms them to a
single ODE for the $z$-component:
\begin{gather}
\Delta ^{\ast 2}w^{\ast }=0,  \label{1.7} \\
z=0;\ z\rightarrow +\infty :\quad w^{\ast }=0.  \label{1.7b}
\end{gather}%
A general solution of the fourth-order ODE (\ref{1.7}) decaying at infinity
can be written as%
\begin{equation*}
w^{\ast }=\left( c_{z}+d_{z}z\right) \exp \left( -\left\vert
k_{n}\right\vert z\right) .
\end{equation*}%
Condition (\ref{1.7b}) then determines the coefficient, $c_{z}=0.$ The
velocity Fourier coefficient in $y$-direction can also be obtained from the
first Eq.~(\ref{2b}):%
\begin{equation}
v^{\ast }=c_{y}\exp \left( -\left\vert k_{n}\right\vert z\right) \left(
1-\left\vert k_{n}\right\vert z\right) ,  \label{1y}
\end{equation}%
with $c_{y}=\mathrm{i}d_{z}/k_{n},$ so that%
\begin{equation}
w^{\ast }=-\mathrm{i}k_{n}c_{y}z\exp \left( -\left\vert k_{n}\right\vert
z\right) .  \label{1z}
\end{equation}%
Thus the solution for a given Fourier mode involves only one unknown
constant $c_{y}\left( n\right) ,\ $ which can be found by applying boundary
conditions (\ref{bcu}). Using (\ref{fu}) and (\ref{1y}), we then get
\begin{equation*}
z=0:\quad v=\sum_{n=-\infty }^{\infty }c_{y}\left( n\right) \exp \left(
\mathrm{i}k_{n}y\right) ,\quad \frac{\partial v}{\partial z}%
=-2\sum_{n=-\infty }^{\infty }\left\vert k_{n}\right\vert c_{y}\left(
n\right) \exp \left( \mathrm{i}k_{n}y\right) ,
\end{equation*}%
so that the slip boundary condition (\ref{bcu}) can be rewritten as
\begin{equation}
c_{y}\left( n\right) +2\sum_{m=-\infty }^{\infty }\left\vert
k_{m}\right\vert c_{y}\left( m\right) b^{\ast }\left( n-m\right) =b^{\ast
}\left( n\right) .  \label{Fby}
\end{equation}

The system (\ref{Fby}) is very similar to that for the longitudinal
patterns, (\ref{Fbx}), and differs only by the prefactor of 2. This means
that the solution for $c_{y}\left( n\right) $ can be expressed in terms of
coefficients $c_{x2}\left( n\right) $ for the longitudinal flow with twice
larger local slip, $b_{2}\left( y\right) =2b\left( y\right) $. Since $%
b_{2}^{\ast }\left( n\right) =2b^{\ast }\left( n\right) $, we obtain from (%
\ref{Fbx}):
\begin{equation}
c_{x2}\left( n\right) +2\sum_{m=-\infty }^{\infty }\left\vert
k_{m}\right\vert c_{x2}\left( m\right) b^{\ast }\left( n-m\right) =2b^{\ast
}\left( n\right) .  \label{fbx2}
\end{equation}%
The left-hand sides of the linear systems (\ref{Fby}) and (\ref{fbx2}) are
identical, but the right-hand sides differ by the factor of 2. In what
follows,%
\begin{equation}
c_{y}\left( n\right) =\frac{c_{x2}\left( n\right) }{2}.  \label{double}
\end{equation}%
Whence by using (\ref{co_b}) we get
\begin{equation}
b_{\mathrm{eff}}^{\bot }\left[ b\left( y\right) /L\right] =\frac{b_{\mathrm{%
eff}}^{\parallel }\left[ 2b\left( y\right) /L\right] }{2}.  \label{aff}
\end{equation}%
Thus, the longitudinal and transverse effective slip lengths are affine,
being related by a simple formula.

\section{Discussion}

In this section we compare Eq.(\ref{aff}) with the results obtained earlier
for some particular periodic textures spatially varying in one direction $y$ along the surface. We
also make use of Eq.(\ref{double}) to prove a similarity of velocity profiles in
eigendirections, again by giving some supporting examples from prior work. Finally, we show that our results are valid for arbitrary, not necessarily periodic, 1D textures (varying on a characteristic scale $L$).

\subsection{Effective slip length}

During the last few decades several theoretical papers have been concerned with the flow past alternating
(parallel or transverse) stripes characterized by piecewise constant,
slip lengths, $b^{+}$ and $b^{-}$ with area fractions $\phi^+$ and $\phi^-=1-\phi^+$, correspondingly. Without loss of generality we consider below that $0 \le b^-\le b^+ < \infty$. A large fraction of these papers deals with an ideal case of stripes with $b^{+} \gg L$ (perfect slip) and $b^{-}=0$ (no-slip). The formula describing the effective slip in longitudinal direction for such a texture was proposed by \cite{philip.jr:1972}:%
\begin{equation}
b_{\mathrm{ideal}}^{\parallel}=\frac{L}{\pi }\ln \left[ \sec \left( \frac{\pi \phi^+
}{2}\right) \right]   \label{Phil72}
\end{equation}%
Later \cite{Lauga:2003} derived an expression for the transverse configuration:%
\begin{equation}
b_{\mathrm{ideal}}^{\perp}=\frac{L}{ 2 \pi }\ln \left[ \sec \left( \frac{\pi \phi^+
}{2}\right) \right],   \label{Lauga}
\end{equation}%
which suggested that transverse and longitudinal components of the slip-length tensor are related as
\begin{equation}
b_{\mathrm{ideal}}^{\perp}= \frac{b_{\mathrm{ideal}}^{\parallel}}{2}.
\label{Phil72b}
\end{equation}%
This relationship is consistent with predictions of Eq.(\ref{aff}). Note that Eq.(\ref{Phil72b}) should be valid not only for regular textures, shown in Fig. \ref{fig:slab}, but also for periodic textures including several stripes of differing widths, e.g., for hierarchical fractal surfaces~\citep{cottin:2012}.

A few authors have discussed a situation of $b^-=0$ and finite $b^+$. Numerical data obtained by \cite{cottin:2004} seem to satisfy Eq.(%
\ref{aff}), and their $b_{\mathrm{eff}}^{\parallel}$, $b_{\mathrm{eff}}^{\perp}$
asymptotically tend to the limiting values, $b_{\mathrm{ideal}}^{\parallel}$ and $b_{\mathrm{ideal}}^{\perp}$,
when $b^+/L$ becomes large.  \cite{Belyaev:2010} derived analytical expressions for this case
\begin{equation*}
b_{\mathrm{eff}}^{||}\simeq \frac{L}{\pi }\frac{\ln \left[ \sec \left( \frac{%
\pi \phi^+ }{2}\right) \right] }{1+\frac{L}{\pi b^+}\ln \left[ \sec \left(
\frac{\pi \phi^+ }{2}\right) +\tan \left( \frac{\pi \phi^+ }{2}\right) \right] },
\end{equation*}%
\begin{equation*}
b_{\mathrm{eff}}^{\bot }\simeq \frac{L}{2\pi }\frac{\ln \left[ \sec \left(
\frac{\pi \phi^+ }{2}\right) \right] }{1+\frac{L}{2\pi b^+ }\ln \left[ \sec
\left( \frac{\pi \phi^+ }{2}\right) +\tan \left( \frac{\pi \phi^+ }{2}\right) %
\right] },
\end{equation*}%
which are again in agreement with predictions of Eq.(\ref{aff}).

\cite{ng:2009} addressed the problem of effective slip lengths for stripes with $b^+\gg L$ and partial $b^-$, and obtained
\begin{equation*}
b_{\mathrm{eff}}^{\parallel}\simeq b_{\mathrm{ideal}}^{\parallel}+\frac{b^-}{\phi^- }%
,\quad b_{\mathrm{eff}}^{\bot }\simeq b_{\mathrm{ideal}}^{\perp}+\frac{b^-%
}{\phi^- }.
\end{equation*}%
It can be seen, that this result also satisfies Eq.(\ref{aff}).

For a weakly slipping anisotropic texture, $b (x,y)\ll L$, the area-averaged
isotropic slip length has been
predicted~\citep{Belyaev:2010}. This means that the slip length tensor
becomes isotropic and for all in-plane directions, the flow aligns with the
applied shear stress. Similar conclusion has been made by \cite%
{Kamrin_etal:2010}. They proposed asymptotic solutions for $\mathbf{b}_{%
\mathrm{eff}}$ for a weakly slipping interface, $\varepsilon =\max
\left\vert b^{\ast }\left( n\right) \right\vert \ll 1.$ The two-term
expansions of the effective slip lengths in $\varepsilon $ can be obtained
for an arbitrary local slip :%
\begin{equation}
b_{\mathrm{eff}}^{||}/L=b^{\ast }\left( 0\right) -\sum_{n=-\infty }^{\infty
}\left\vert k_{n}\right\vert \left\vert b^{\ast }\left( n\right) \right\vert
^{2},\quad b_{\mathrm{eff}}^{\bot }/L=b^{\ast }\left( 0\right)
-2\sum_{n=-\infty }^{\infty }\left\vert k_{n}\right\vert \left\vert b^{\ast
}\left( n\right) \right\vert ^{2}.  \label{Kamr10}
\end{equation}%
Since $b_{2}^{\ast }\left( n\right) =2b^{\ast }\left( n\right) $, Eq.(\ref%
{Kamr10}) is also consistent with the relation (\ref{aff}).

These examples fully support our predictions, but of course Eq.(\ref{aff})
is universal and should hold for any 1D surface.

Moreover, Eq.(\ref{aff}) still holds when the flow is unsteady. \cite{ng:2011}
have considered pressure-driven oscillatory flow in a channel with walls
patterned by stripes of $b^+\gg L$ and $b^-=0$. For a thick channel they found that both real and imaginary parts of the effective slip lengths roughly satisfy Eq.(\ref{Phil72b}). The reason is that
Eqs.(\ref{1x}), (\ref{Fbx}) and (\ref{1y}-\ref{Fby}) remain valid in this
case if one replaces $\left\vert k_{n}\right\vert $ by $\sqrt{k_{n}^{2}+%
\mathrm{i}\omega }.$

\subsection{Flow field}

We remark and stress that Eq.(\ref{double}) allows one to express the entire
flow field for the transverse configuration of patterns in terms of the
longitudinal flow field, $u_{2}\left( y,z\right) =u\left[ y,z,2b\left(
y\right) \right] .\ $ Indeed, by applying the inverse Fourier transform of (%
\ref{1y}) and (\ref{1z}), one can easily derive by using (\ref{1x}), (\ref%
{2b}) and (\ref{double}):%
\begin{gather}
v=\frac{1}{2}\left( u_{2}+z\frac{\partial u_{2}}{\partial z}\right) ,\quad
w=-\frac{z}{2}\frac{\partial u_{2}}{\partial y},  \label{vw} \\
p=-\frac{\partial u_{2}}{\partial y}.  \label{p}
\end{gather}%
Whence we conclude that at the wall%
\begin{equation}
z=0:\quad v=\frac{1}{2}u_{2},\quad \frac{\partial v}{\partial z}=\frac{%
\partial u_{2}}{\partial z}.  \label{v0}
\end{equation}

\begin{figure}[tbp]
\begin{center}
\vspace{0.1in} \includegraphics[width=3.5in]{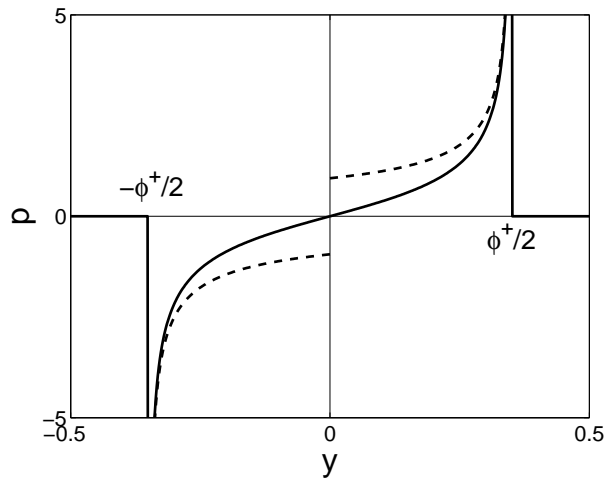}
\end{center}
\caption{Pressure distribution in the transverse flow over alternating
perfect-slip, $\left\vert y\right\vert \leq \phi ^{+}/2$, and no-slip, $%
\phi ^{+}/2\geq \left\vert y\right\vert \geq 1/2,$ stripes. Solid and dashed
lines are Eqs. (\protect\ref{p_ph}) and (\protect\ref{p_as}), respectively.}
\label{fig:p}
\end{figure}

Several important conclusions follow from (\ref{vw}), (\ref{p}) for
alternating perfect slip ($b^{+}\gg L $) and no-slip stripes $\left(
b^{-}=0\right) .$ In this case $u_{2}=u,$ and hence, at the wall the
velocity along this 1D texture is always twice that of perpendicular to it, $%
u=2v$. The same relation between velocities in eigendirections has
been found by \cite{Teo&Khoo:09} and \cite{ng:2010} for the pressure-driven
flow in a wide grooved superhydrophobic channel. Analytical solutions have been
obtained for the longitudinal velocity $u$ and its gradient $\partial
u/\partial y$ at the wall for the shear flow \citep{sbragaglia.m:2007}. Eq.(%
\ref{p}) enables us to obtain the pressure for the transverse flow over the
perfect-slip region, $\left\vert y\right\vert \leq \phi ^{+}/2$:%
\begin{equation}
z=0:\quad p=-\frac{\partial u}{\partial y}=\frac{\sin \left( \pi y\right) }{%
\sqrt{\cos ^{2}\left( \pi y\right) -\cos ^{2}\left( \pi \phi ^{+}/2\right) }}%
.  \label{p_ph}
\end{equation}%
The pressure over the no-slip regions, $\phi ^{+}/2\geq \left\vert
y\right\vert \geq 1/2,$ where $\partial u/\partial y=0,$ is zero. The
distribution (\ref{p_ph}) calculated for $\phi ^{+}=0.7$ is shown in Fig. %
\ref{fig:p}. It grows infinitely near the jump in $b(y)$ (from $b^{+}\gg L$
to $b^{-}=0$) at $y=\pm \phi ^{+}/2$:%
\begin{equation}
p\simeq \pm \sin \left( \pi \phi ^{+}/2\right) \left[ \pi \sin \left( \pi
\phi ^{+}\right) s\right] ^{-1/2}\quad \text{as}\quad s=\left( \phi
^{+}/2-\left\vert y\right\vert \right) \rightarrow +0.  \label{p_as}
\end{equation}

Note that a solution for a velocity in an arbitrary direction of the
undisturbed flow, $\mathbf{U}=z\left( \mathbf{e}_{x}\cos \theta +\mathbf{e}%
_{y}\sin \theta \right) \mathbf{,}$ represents a superposition of solutions
in eigendirections:%
\begin{equation*}
\mathbf{u}_{1}=u\mathbf{e}_{x}\cos \theta +\left( v\mathbf{e}_{y}+w\mathbf{e}%
_{z}\right) \sin \theta .
\end{equation*}%
Therefore, the knowledge about the longitudinal velocity at the wall allows
us to calculate the flow field in any direction given by an angle $\theta $.

\subsection{Surfaces with an arbitrary 1D texture}

Up to this point, our focus has been on periodic 1D surface. Now,
we can remove all periodicity requirements we used to derive Eq.(%
\ref{double}) by applying the Fourier analysis. Indeed, if $u_{2}$ is the
solution of the longitudinal problem with a non-periodic slip length $2L\beta
\left( y\right) $, satisfying
\begin{equation*}
\Delta u_{2}=0,
\end{equation*}%
\begin{equation*}
z=0:\quad u_{2}-2\beta \left( y\right) \frac{\partial u_{2}}{\partial z}%
=2\beta \left( y\right) ,
\end{equation*}%
then one can verify directly that transverse solutions, Eqs.(\ref{vw}), (\ref%
{p}), taking into account (\ref{v0}), satisfy the Stokes equation (\ref{Se}) and boundary conditions (\ref{bcu}), (\ref{bcw0}). Thus, our results are valid for any arbitrary patterned, not necessarily periodic, 1D surfaces.

\section*{Acknowledgement}

This research was supported by the Russian Academy of Sciences through its
Priority Programme `Assembly and Investigation of Macromolecular Structures
of New Generations'
\bibliographystyle{jfm}
\bibliography{slipsphere}

\end{document}